\documentstyle[preprint,revtex]{aps}                  
\begin{document}
\draft
\pagestyle{empty}                                     
\centerline{\hfill                 NTUTH--94-21} 
\centerline{\hfill                 December 1994} 
\vfill                                            
\begin{title}
Perspective on Quark Mass and Mixing Relations
\end{title}
\vfill                                            
\author{Wei-Shu Hou and Gwo-Guang Wong}
%
\begin{instit}
Department of Physics,
National Taiwan University,
Taipei, Taiwan 10764, R.O.C.
\end{instit}
\receipt{\today}
\vfill                             
\begin{abstract}
Recent data indicate that
$V_{ub}\cong \lambda^4 \cong (0.22)^4$,
while $m_t$ seems to be $174$ GeV.
The relations
$m_d/m_s\sim m_s/m_b \sim \delta \sim \lambda^2 \simeq \vert V_{cb}\vert$
and
$m_u/m_c\sim m_c/m_t \sim \delta^2 \sim \lambda^4 \sim \vert V_{ub}\vert$
suggest that
the down type sector is responsible for $\vert V_{us}\vert$
and $\vert V_{cb}\vert$,
while $V_{ub}$ comes from the up type sector.
Five to six parameters might suffice
to account for the ten quark mass and mixing parameters,
resulting in specific power series representations for the mass matrices.
In this picture, $\delta$ seems to be the more sensible expansion parameter,
while $\lambda \cong \sqrt{m_d/m_s} \sim \sqrt{\delta}$
is tied empirically to  $(M_d)_{11} = 0$.
\end{abstract}

\vfill                             
\narrowtext
\def\ltap{\ \raisebox{-.5ex}{\rlap{$\sim$}} \raisebox{.4ex}{$<$}\ }
\def\gtap{\ \raisebox{-.5ex}{\rlap{$\sim$}} \raisebox{.4ex}{$>$}\ }
\pagestyle{plain}


Eleven years ago,
when the prolonged $B$ lifetime became known,
Wolfenstein suggested \cite{Wolf} that
$\lambda \equiv V_{us} \simeq 0.22$ could be
an expansion parameter for
the Kobayashi-Maskawa (KM) mixing matrix \cite{KM}.
At that time, the experimental values were
\begin{equation}
V_{cb} \approx 0.06,\ \ \ \ \vert V_{ub}/V_{cb} \vert < 0.2,
\label{eq:old}
\end{equation}
hence Wolfenstein proposed to parametrize the KM matrix as
\begin{equation}
V= \left[ \begin{array}{ccc}
          \ 1-{1\over 2}\lambda^2 \ & \lambda \ & \
          A\lambda^3(\rho-i\eta) \\
          \ -\lambda\ & \
          \ 1-{1\over 2}\lambda^2 \ & \
          \ A\lambda^2 \\
          \ A\lambda^3(1-\rho-i\eta) \ & \ -A\lambda^2 \ & \ 1
          \end{array} \right],
\label{eq:Wolf}
\end{equation}
to order $\lambda^3$, with $A \approx 5/4$ and $\rho^2 + \eta^2 <1$.

The parametrization has since become a reference standard \cite{PDG},
especially for $CP$ violation studies.
In the past few years, however,
the experimental values for $V_{cb}$ and $\vert V_{ub}/V_{cb} \vert$
have been consistently dropping \cite{PDG}.
The current values are
\begin{equation}
\vert V_{cb} \vert = 0.040 \pm 0.005,
                                                                         \ \ \
\
\vert V_{ub}/V_{cb} \vert = 0.08 \pm 0.02.
\label{eq:new}
\end{equation}
Thus, $A$ is now $0.8 \pm 0.1$,
which is down by 1/3 compared to ten years ago.
Present trends in both theory and experiment suggest \cite{ICHEP}
that $V_{cb}$ may drop a little further, to below $0.04$.
More dramatic is the factor of 4
drop in $\vert V_{ub}\vert$ from that of eq. (\ref{eq:old}),
\begin{equation}
\vert V_{ub}\vert = 0.0032 \pm 0.0009.
\label{eq:Vub}
\end{equation}
Noting that $V_{us} \cong 0.2205$
hence $\lambda^2 \cong 0.0486$, $\lambda^3 \cong 0.0107$
and $\lambda^4 \simeq 0.0024$,
we set
\begin{equation}
V_{ub} \equiv A\lambda^4(\rho^\prime - i \eta^\prime) \equiv B\lambda^4
e^{-i\phi},
\label{eq:rho'}
\end{equation}
where $\rho \equiv \rho^\prime\lambda$, $\eta \equiv \eta^\prime\lambda$
(i.e. $d\rho/d\lambda = \rho^\prime$), and, numerically,
\begin{equation}
B \equiv A\sqrt{\rho^{\prime 2} + \eta^{\prime 2}} = 1.3 \pm 0.5.
\label{eq:num}
\end{equation}

At first sight, this may seem to be mere numerology.
However, in the spirit of Wolfenstein's original proposal \cite{Wolf},
a change in order of $\lambda$ may have
profound implications for possible underlying dynamics
that could relate the mixing angles and quark mass ratios.
Note that the recent CDF result  \cite{CDF}
of $m_t(m_t)  \cong 174$ GeV,
gives $m_t(1$ GeV$) \sim 360$ GeV,
hence $m_c/m_t \sim 1/280$.
With the values $m_d/m_s \simeq 1/21 - 1/18$,
$m_s/m_b \simeq 1/40 - 1/25$ and
$m_u/m_c \simeq 1/390 -1/200$ \cite{GL},
we now seem to have
\begin{equation}
          m_d/m_s  \sim   m_s/m_b \sim \lambda^2
                                               \gg
          m_u/m_c  \sim   m_c/m_t \sim \lambda^4.
\label{eq:mratio}
\end{equation}
In relation to this, several authors have noted \cite{sob,Ng} that
the relation $V_{cb} \simeq m_s/m_b$ now holds rather well.
Together with the old empirical relation
\begin{equation}
V_{us} = \lambda \simeq \sqrt{m_d/m_s},
\label{eq:Wein}
\end{equation}
it appears that the KM matrix $V$ is
mostly due to the down type sector, while
eqs. (\ref{eq:rho'}) and (\ref{eq:mratio}) suggest that perhaps $V_{ub}$
originates from the up type sector.
We wish to explore this proposition, emphasizing
that the main ``Gestalt" switch stems from the change
in $V_{ub}:\ \lambda^3 \longrightarrow \lambda^4$.


Within the Standard Model (SM), we can always redefine
the right-handed quark fields to make the mass matrices hermitian.
Although this limits the
form of the possible underlying dynamics,
we do so for simplicity.
Ignoring the $u$-type quark mass matrix to first approximation
since its mass ratios are subdominant,
we have
\begin{equation}
M_d \cong V \left[ \begin{array}{ccc}
                        \ -m_d \ & \ 0 \ & \ 0 \\
                        \ 0 \ & \ m_s \ & \ 0 \\
                        \ 0 \ & \ 0 \ & \ m_b \\
          \end{array} \right] V^\dagger.
\label{eq:Ma}
\end{equation}

It is desirable to maintain the empirical relation of eq. (\ref{eq:Wein}).
As Ma has argued \cite{Ma},
this is guaranteed if $(M_d)_{11} = 0$ in eq. (\ref{eq:Ma}).
It is approximately true numerically, namely
$(M_d)_{11} = - m_d \vert V_{ud}\vert^2
                          + m_s \vert V_{us}\vert^2 + m_b \vert V_{ub}\vert^2
\cong 0$,
as can be most easily checked by making a series expansion in $\lambda$.
Arguing that perhaps $(M_dM_d^\dagger)_{12} \simeq 0$,
Ma suggested a second relation \cite{Ma},
$m_s^2/m_b^2 = -V_{ub} V_{cb}/V_{us}$.
With Wolfenstein's assignment that $V_{ub} \sim \lambda^3$,
and experimental values for $V_{ub}$ and $V_{cb}$ up till a few years ago,
this relation seemed plausible.
With eq. (\ref{eq:new}), however, the relation no longer appears to hold,
as can be most easily checked again via a series expansion  in $\lambda$.
Thus, the discrete symmetry $S_3\times Z_3$
proposed by Ma is no longer well motivated.
Instead, as mentioned earlier, the recently popular assignment is \cite{sob,Ng}
\begin{equation}
m_s/m_b \cong V_{cb} \equiv A\lambda^2,
\label{eq:sob}
\end{equation}
which becomes even more appealing if $V_{cb}$ drops below 0.04.
Eqs. (\ref{eq:rho'}) and (\ref{eq:mratio})
suggest that one may relegate the generation of $V_{ub}$,
hence $CP$ violation, to the $u$-type quark sector.
With this in mind, without loss of generality,
we redefine quark fields to make $M_d$ real symmetric.
We ask ourselves what is the {\it least number of parameters}
needed to account for both $d$-type quark masses and
$V$, with $V_{ub} = 0$.
Thus, $D_L \equiv \left. V\right|_{V_{ub} = 0}$,
and has just two parameters $\lambda$ and $A$,
that is
\begin{equation}
D_L = \left[ \begin{array}{ccc}
          \ \sqrt{1-\lambda^2}    &    \lambda    &    \ 0 \\
          \ -\lambda \sqrt{1-\delta^2} \  &
          \ \sqrt{1-\lambda^2} \sqrt{1-\delta^2} & \ \delta \\
          \ \lambda\delta    &    \ - \sqrt{1-\lambda^2}\, \delta
                                      &    \ \sqrt{1-\delta^2}
          \end{array} \right],
\label{eq:DL}
\end{equation}
where $\delta \equiv A\lambda^2$.
Strictly speaking, $\lambda$ and $\delta$ of eq. (\ref{eq:DL})
should be $s_{12}$, $s_{23}$
of standard convention \cite{PDG}.
However, they are basically just Wolfenstein's $\lambda$ and
$A\lambda^2$ up to $c_{13} =1 + O(\lambda^8)$
correction factor.
Ma's observation is now reformulated as
\begin{equation}
(M_d)_{11} = - m_d (D_L)_{11}^2
                          + m_s (D_L)_{12}^2 = 0.
\label{eq:Mad}
\end{equation}
The
strict (``texture") zero implies the relation 
\begin{equation}
 V_{us} \equiv \lambda  \cong s_{12} = \sqrt{m_d\over m_s + m_d},
\label{eq:lambda}
\end{equation}
or $\tan \theta_{12} = \sqrt{m_d/m_s}$.
To order $\lambda^4$,
eqs. (\ref{eq:sob}) and (\ref{eq:lambda}) together imply that,
\begin{equation}
(-\hat m_d,\ \hat m_s,\ \hat m_b) = (-A\lambda^4,\ A\lambda^2 - A\lambda^4, 1)
,
\label{eq:Mdeigen}
\end{equation}
where we normalize to $m_b$.
Multiplying
$\hat m_d$ and $\hat m_s$ by a function $g(\lambda)$
with  $g(0) = 1$ does not alter $\hat m_d$ at $\lambda^4$ order,
but modifies $\hat m_s$ at $\lambda^3$ order. 
To reduce this ambiguity,
we suppress the odd powers in $\lambda$,
and set $g^{\prime\prime}(0) = 0$.
Hence, $g(\lambda) = 1 + O(\lambda^4)$, suggesting
that perhaps $g(\lambda) = 1$ to all orders.
Multiplying out eq.  (\ref{eq:Ma}),
we find 
\begin{equation}
\hat M_d = \left[ \begin{array}{ccc}
             \ 0 & \ \ A\lambda^3 & \ \ 0 \\
             \ A\lambda^3 & \ \ A\lambda^2+(A-2)A\lambda^4 & \
             \ A\lambda^2-A^2 \lambda^4 \\
             \ 0 & \ \ A\lambda^2-A^2\lambda^4 & \ \ 1 - A^2\lambda^4
             \end{array} \right],
\label{eq:Md}
\end{equation}
at least to $\lambda^4$ order.
Thus, with  $m_b$, $A$ ($\sim 1$) and $\lambda$, one can
account for the 5 parameters $m_d$, $m_s$, $m_b$ and
$\vert V_{us}\vert$, $\vert V_{cb}\vert$, resulting in two relations,
eq. (\ref{eq:lambda}), and
\begin{equation}
V_{cb} \equiv A\lambda^2 \cong s_{23} = {m_s + m_d\over m_b},
\label{eq:Al2}
\end{equation}
which are slight modifications of
eqs. (\ref{eq:Wein}) and (\ref{eq:sob}).


{\it Postulating} that $D_L$ of eq. (\ref{eq:DL}) accounts for
$V$ up to $V_{ub} = 0$, we turn to the up type quark sector.
The diagonalization matrix $U_L \equiv D_L V^\dagger$ is
\begin{equation}
U_L
     \simeq \left[ \begin{array}{ccc}
         \ 1 \   &   \ 0 \   &  \ -B \lambda^4 e^{-i\phi} \  \\
           \ 0 \   &   \ 1 \   &   \ 0 \ \\
           \ B \lambda^4 e^{i\phi} \ & \ 0 \ & \ 1 \
          \end{array} \right],
\label{eq:UL}
\end{equation}
up to corrections of order $\lambda^6$,
and
$M_u = U_L\, \mbox{diag}\, (-m_u,\ m_c,\ m_t)\, U_L^\dagger$.
Noting from eq. (\ref{eq:mratio}) that
$m_t : m_c : m_u \approx 1 : \lambda^4 : \lambda^8$,
we find that, up to order $\lambda^8$ corrections,
\begin{equation}
\hat M_u \cong
          \left[ \begin{array}{ccc}
          \ 0 & \ 0 & \ -B \lambda^4 e^{-i\phi} \\
          \ 0 & \ C\lambda^4 & \ 0 \\
          \ -B \lambda^4 e^{i\phi} & \ 0 & \ 1
          \end{array} \right],
\label{eq:Mu}
\end{equation}
where we normalize to $m_t$.
In fact, for the zeros of $(\hat M_u)_{ij}$, $i+j = odd$,
the corrections
are at order $\lambda^{10}$.
The order $\lambda^8$ correction to $(\hat M_u)_{11}$
is removed by the condition
$(\hat M_u)_{11} = - m_u \vert U_L\vert_{11}^2
                          + m_t \vert U_L\vert_{13}^2 = 0$
up to order $\lambda^{12}$,
which is similar to but weaker than eq. (\ref{eq:Mad}),
resulting in the relation
\begin{equation}
\vert V_{ub}\vert \equiv B\lambda^4 \cong \sqrt{m_u\over m_t}.
\label{eq:Bl4}
\end{equation}
Thus both $V_{ub}$ and $m_u$ are generated via
diagonalizing the $u$-$t$ mixing element.
The number of parameters
are further reduced if
$C = B$, leading to a second relation
\begin{equation}
m_u/m_c=m_c/m_t\ (=B\lambda^4),
\label{eq:geo}
\end{equation}
which we call the geometric relation.
Since $B \simeq 1$ (eq. (\ref{eq:num})),
it may well be that $B = 1$ or $1/A$ in Nature,
such that
$m_u/m_c = m_c/m_t = \vert V_{ub}\vert = \lambda^4$
or $A^{-1}\lambda^4$,
and just {\it two} additional parameters,
$m_t$ and $\phi$,
{\it might} account for the remaining {\it five} parameters
$m_u$, $m_c$, $m_t$ and $V_{ub} = \vert V_{ub}\vert e^{-i\phi}$.
Given the uncertainties in mixing,
and especially in the {\it lighter} quark masses \cite{PDG},
these reults are not inconsistent with data!

It is clear how these relations can be weakened.
Choosing to maintain eq. (\ref{eq:Md}),
the corrections are relegated to eq. (\ref{eq:Mu}).
First,
restoring $C\neq 1$ one could fine tune $m_c$.
Second,
$B\neq C$ breaks the relation of eq. (\ref{eq:geo}).
Third,
$(\hat M_u)_{11}$ can be nonvanishing at
order $\lambda^8$, breaking the relation of eq. (\ref{eq:Bl4}).
Finally, two parameters may be introduced at
order $\lambda^3$ and $\lambda^6$
to $(\hat M_u)_{23}$ and $(\hat M_u)_{12}$,
as order $\lambda^3$ and $\lambda^2$ corrections to
$V_{cb}$ and $V_{us}$, respectively.
At this stage, however, although approximate mass--mixing relations
still hold, one has recovered the full set of 10 parameters.
Furthermore, with  $(\hat M_u)_{23}$ or $(\hat M_u)_{12}$ restored,
in principle one needs to introduce a second phase, such that
a phase redefinition is necessary to get back to
the standard phase convention.

Since we follow Wolfenstein in
expanding $V$ in powers of $\lambda$,
the phenomenological consequences are
the same as the Standard Model.
In particular, large $m_t$
is balanced by the smallness of  $V_{ub}$
hence $\varepsilon_K$ is accountable,
while it is known theoretically that
$\varepsilon^\prime/\varepsilon$ is of order $10^{-3}$
but may well be vanishingly small \cite{Buras}.
Hence, it could be consistent with
either E731 or NA31 values \cite{PDG}.
$B_d$ mixing is consistent with $m_t \cong 174$ GeV
if one takes into account the uncertainties in $f_B\sqrt{B_B}$.
The $CP$ violating invariant
$J_{CP}$ is of order $\lambda^7$:
$
J_{CP} = \mbox{\rm Im}(V_{us} V_{ub}^\ast V_{cs}^\ast V_{cb})
\simeq A^2 \eta^\prime \lambda^7 \ltap 3\times 10^{-5}.
$ 
For the unitarity triangle, one can continue to use $\rho$ and
$\eta$ of Wolfenstein \cite{PDG}.
However, given that
$\rho$, $\eta = \rho^\prime\lambda$, $\eta^\prime\lambda$
with $\sqrt{\rho^{\prime 2} + \eta^{\prime 2}} \sim 1.6$,
the unitarity triangle appears a bit squashed.
The phase angle $\phi \equiv \tan^{-1} \eta^\prime/\rho^\prime$
is still a free parameter.


Our results are similar to the ansatz of Ng and Ng \cite{Ng},
where the starting point is also the approximate relations of
eq. (\ref{eq:mratio}). It is also close to that of Giudice \cite{Giudice},
where the setting is supersymmetric Grand Unified Theories (SUSY GUTS).
In both cases the ansatz appears without much justification.
{}From the perspective of Wolfenstein's $\lambda$ expansion, however,
we make the key observation that $V_{ub}$ is of  $\lambda^4$,
eq. (\ref{eq:rho'}), which is more consistent with data.
Together with eq. (\ref{eq:mratio}),
it suggests that $V_{us}$ and $V_{cb}$ originate from $M_d$
while  $V_{ub}$ arises from $M_u$.
Demanding for {\it least number of parameters},
which is related in spirit but not equivalent to finding ``texture zeros"
\cite{RRR},
we ``deduce" the mass matrices of eqs. (\ref{eq:Md}) and (\ref{eq:Mu}),
by idealizing the input equations
(\ref{eq:rho'}), (\ref{eq:mratio}), (\ref{eq:Wein}) and (\ref{eq:sob}).
Note that Ng and Ng conclude that $(\hat M_d)_{32} = 0$,
which is not necessary, while Giudice arbitrarily
sets $(M_d)_{23} = 2\,(M_d)_{22}$.

It would be appealing if some symmetry
or dynamical mechanism underlies the
possible reduction of 2 to 5 parameters
from the 10 quark masses and mixing angles.
Discrete symmetries \` a la Ma \cite{Ma}
can be found, for example $Z_8\times Z_2$
with 5--6 additional Higgs doublets.
However,
these usually do not add insight to the $\lambda^n$ power behavior
for mixing angles and mass ratios.
Note that eqs. (\ref{eq:Md}) and (\ref{eq:Mu}) suggest
an expansion in even powers of $\lambda$,
{\it except} for $(\hat M_d)_{12} = (\hat M_d)_{21} \cong \lambda^3$.
This is traced to the fact that,
after changing $V_{ub}$ from order $\lambda^3$
to order $\lambda^4$,
the only term odd in $\lambda$ is just
$\vert V_{us}\vert = \vert V_{cd}\vert$ itself.
Defining as before $\delta \equiv V_{cb} \equiv A\lambda^2$,
we find, to leading order in $\delta$,
\begin{equation}
\hat M_d \cong \left[ \begin{array}{ccc}
             \ 0 & \ \ \lambda\,\delta & \ \ 0 \\
             \ \lambda\,\delta & \ \ \delta & \ \ \delta \\
             \ 0 & \ \ \delta & \ \ 1
             \end{array} \right], \ \ \ \
\hat M_u \simeq
          \left[ \begin{array}{ccc}
          \ 0 & \ 0 & \ -\delta^2 e^{-i\phi} \\
          \ 0 & \ \delta^2 & \ 0 \\
          \ -\delta^2 e^{i\phi} & \ 0 & \ 1
          \end{array} \right].
\label{eq:ansatz}
\end{equation}
It seems that $\delta \simeq 1/20 - 1/30$ is
the actual expansion parameter,
while Wolfenstein's $\lambda = \sqrt{\delta/A}$
is more puzzling, even though it is empirically tied to
$(\hat M_d)_{11} = 0$ \cite{Ma}.
We find eq. (\ref{eq:ansatz}) to be suggestive of an underlying
radiative mechanism, perhaps not far above the electroweak scale \cite{radm}.
It need not have a high scale origin
such as from SUSY and/or GUTS \cite{RRR}.
We believe that the mass and mixing hierarchies,
with correlations as exemplified in eq. (\ref{eq:ansatz}),
cannot be just an accident.

We offer some final remarks in passing.
Eqs. (\ref{eq:rho'}), (\ref{eq:mratio}), (\ref{eq:Wein}) and (\ref{eq:sob})
imply that
$m_d/m_b = A\lambda^4 \ltap m_u/m_c \sim m_c/m_t
 \sim \vert V_{ub}\vert \sim \lambda^4$,
so $V_{ub}$ may arise from $\hat M_d$
with $\hat M_u \simeq \mbox{diag}\,(-\lambda^8,\ \lambda^4,\ 1)$.
This corresponds to taking the off-diagonal piece from $\hat M_u$
in eq. (\ref{eq:ansatz}) and placing it in $\hat M_d$.
While possible, we find this lacking in appeal as compared
to eq. (\ref{eq:ansatz}), where we
{\it couple the smallness of $CP$ violation effects to the existence and
heaviness of the top quark},
which seems more natural.
The placement of $CP$ phase in $(M_u)_{13}$ may
therefore be rooted in dynamics.
Second,
from eqs. (\ref{eq:lambda}) and (\ref{eq:Al2}),
$A\, m_d m_b = (m_s + m_d)^2$,
hence the $d$-type quarks have a modified geometric relation.
Perhaps eq. (\ref{eq:geo}) should be modified accordingly.
Third,
although we choose $m_t$, $m_b$, 
$\lambda$, $A$ (or, $\delta$ and $A$), 
$\phi$ and possibly $B$ to account for
10 quark mass and mixing parameters,
we note that $A \sim 1$ (and $B\sim 1$)
while the smallness of $m_b/m_t$ is not explained.
Since  $m_b/m_t = (m_b/m_c)(m_c/m_t)
\sim (m_b/m_c)\lambda^4 \sim A\lambda^3$,
perhaps both $A$ and the odd power $\lambda$
expansion are related to $m_b$ generation from a heavy top.

In summary,
recent values for $V_{ub}$, $V_{cb}$ and $m_t$
suggest that the KM matrix originates from the
down type quark sector, except that $V_{ub}$,
as the source of $CP$ violation,
may be due to the up type quark sector.
Following Wolfenstein, we expand
$V$ in terms of a small parameter, but change the order of
$V_{ub}$ from $\lambda^3$ to $\lambda^4$ (eq. (\ref{eq:rho'})).
We construct explicitly
$M_d$ and $M_u$ as given in eqs. (\ref{eq:Md})
and (\ref{eq:Mu})
(or more generically in eq. (\ref{eq:ansatz}),
with $\delta \equiv V_{cb} \sim  \lambda^2$ as expansion parameter).
Just 5 (or 6) parameters
$m_t$, $m_b$, $\lambda$, $A\sim 1$ and $CP$ violating phase $\phi$
(and $B\sim 1$) seem sufficient to account for 10 quark mass and mixing
parameters.
Eqs. (\ref{eq:lambda}) and (\ref{eq:Al2}) extend
the old(er) relations of eqs. (\ref{eq:Wein}) and (\ref{eq:sob}),
while eqs. (\ref{eq:Bl4}) and (\ref{eq:geo})
relate $V_{ub}$ and up type quark mass ratios.
These relations may serve as starting points for possible small corrections.

\acknowledgments
We thank E. Ma and D. Ng for useful discussions.
The work of WSH is supported in part by grants NSC 84-2112-M-002-011,
and GGW by NSC 84-2811-M-002-035
of the Republic of China.

\end{document}